%% file: IEEE-conference-template-062824.tex
\documentclass[conference]{IEEEtran}
\IEEEoverridecommandlockouts

\usepackage{cite}
\usepackage{booktabs}
\usepackage{amsmath,amssymb,amsfonts}
\usepackage{graphicx}
\usepackage{textcomp}
\usepackage{xcolor}
\def\BibTeX{{\rm B\kern-.05em{\sc i\kern-.025em b}\kern-.08em
    T\kern-.1667em\lower.7ex\hbox{E}\kern-.125emX}}

\usepackage{comment}
\usepackage[utf8]{inputenc}
\usepackage{amsmath,amssymb,amsfonts}

\usepackage{makecell}
\usepackage{array}
\usepackage{multirow}
\usepackage{graphicx}  
\usepackage{url}
\usepackage{subcaption}
\usepackage{amsmath}
\usepackage{amssymb}
\usepackage{algorithm}
\usepackage{algpseudocode}
\usepackage{arydshln} 
\usepackage{adjustbox}

\usepackage{pgfplots}
\pgfplotsset{compat=1.18}
\usepackage{caption}
\usepackage{subcaption}
\usepackage{graphicx}

\definecolor{dark_green}{RGB}{52, 83, 67}       
\definecolor{light_green}{RGB}{142, 190, 71}  
\usepackage{xcolor}
\definecolor{fake_color}{RGB}{211,128,135}
\definecolor{real_color}{RGB}{101,139,188}

\begin{document}


\title{Fake-Mamba: Real-Time Speech Deepfake Detection Using Bidirectional Mamba as Self-Attention's Alternative}

\author{
Xi Xuan\textsuperscript{1,*}, 
Zimo Zhu\textsuperscript{2}, 
Wenxin Zhang\textsuperscript{3,4}, 
Yi-Cheng Lin\textsuperscript{5}, 
Tomi Kinnunen\textsuperscript{1}\\
\IEEEauthorblockA{
\textsuperscript{1}School of Computing, University of Eastern Finland, Finland \\
\textsuperscript{2}Department of Statistics and Applied Probability, University of California Santa Barbara, USA \\
\textsuperscript{3}School of Computer Science and Technology, University of Chinese Academy of Sciences, China \\
\textsuperscript{4}Department of Mathematics, University of Toronto, Canada \\
\textsuperscript{5}Graduate Institute of Communication Engineering, National Taiwan University, Taiwan \\
*Corresponding author: xi.xuan@uef.fi
}
}

\maketitle

\begin{abstract}
Advances in speech synthesis intensify security threats, motivating real-time deepfake detection research. We investigate whether bidirectional Mamba can serve as a competitive alternative to Self-Attention in detecting synthetic speech. Our solution, Fake-Mamba, integrates an XLSR front-end with bidirectional Mamba to capture both local and global artifacts. Our core innovation introduces three efficient encoders: TransBiMamba, ConBiMamba, and PN-BiMamba. Leveraging XLSR's rich linguistic representations, PN-BiMamba can effectively capture the subtle cues of synthetic speech. Evaluated on ASVspoof 21 LA, 21 DF, and In-The-Wild benchmarks, Fake-Mamba achieves 0.97 \%, 1.74 \%, and 5.85 \% EER, respectively, representing substantial relative gains over SOTA models XLSR-Conformer and XLSR-Mamba. The framework maintains real-time inference across utterance lengths, demonstrating strong generalization and practical viability. The code is available at {\url{https://github.com/xuanxixi/Fake-Mamba}}.
\end{abstract}


\section{Introduction}

Powered by advanced deep generative neural networks, recent \emph{text-to-speech} (TTS)~\cite{ref1, breezyvoice} and \emph{voice conversion} (VC)~\cite{ref2} systems can generate highly realistic artificial or modified speech. TTS and VC mainly use in assistive technology, games, user interfaces,  and audiobooks~\cite{ref3}, they also pose many risks, ranging from legal perjury~\cite{Xuan2025}, financial fraud~\cite{zhang1}, and political discord~\cite{ding1} to spoofing voice biometric systems. Consequently, speech deepfake detection (SDD) has emerged as an active area of research~\cite{ 10832250, codecfake-omni}. 

SDD can be formulated as a sequence classification task. Its objective is to learn discriminative latent features and temporal dependencies within entire utterances, utilizing effective encoding schemes to classify each utterance as either human or synthetic speech. With its hybrid architecture, Conformer~\cite{ref21,xuan2024conformer} excels in SDD tasks by combining CNN and Transformer to extract global and local features, leveraging self-attention to capture long-term dependencies like intonation and semantic context for more comprehensive speech modeling. Many studies explore the extension of the Conformer-based methods to enhance their performance in SDD~\cite{ref17,ref47}.

Although these models perform well in SDD, their core component, Conformer, has several limitations. We identify several key issues with the current SOTA SDD methods:

\begin{itemize}
    \item \textbf{Quadratic time complexity of MHSA~\cite{vaswani2017attention}}. The time complexity of MHSA in the Conformer scales as $\mathcal{O}(t^2)$, where $t$ is the sequence length. This imposes limitations specifically on memory-limited edge devices. 

    \item \textbf{Robustness and generalization issues in Conformer~\cite{ref21}.} 
    Even if synthetic speech artifacts appear to be distributed across specific time-frequency regions 
    \cite{ref22, ref23,zhang2,zhang3,zhang4}, the \emph{multi-head self-attention} (MHSA) mechanism \cite{vaswani2017attention} utilizes the dot product between input tokens along the temporal dimension, 
    potentially overlooking the dependencies between the temporal and channel dimensions~\cite{ref47,linprime}. This may compromise robustness to speech coding \cite{wu-etal-2024-codec, 10849259} (relevant to call-center applications) and audio compression (relevant to social media platforms) that are unavoidable in applications. 
\end{itemize}

To address these limitations, and to explore more efficient and novel architectures for SDD, we explore the potential of Mamba, a state-space model~\cite{ref26}. Mamba has recently achieved state-of-the-art performance across various domains, including language modeling~\cite{zhao2025cobra,lenz2025jamba}, computer vision \cite{ref28, ref29} and time-series modeling \cite{wang2025mamba,li2024cmmamba}. Crucially, Mamba offers two compelling advantages over Conformer-based approaches: (1) near-linear time complexity and (2) a global receptive field.  
Unlike Conformer networks, which rely on MHSA for temporal token selection, Mamba’s input-dependent selection mechanism enables more efficient information flow by dynamically controlling feature contributions to hidden states. This allows the model to minimize irrelevant influences during state computation across both time and feature channels. Such implicit feature selection enhances crucial artifact detection capabilities for speech deepfakes while significantly reducing computational overhead. Leveraging these strengths, we investigate Mamba’s capability to localize subtle artifacts, further enhancing deepfake speech detection. 

Building on these insights, we propose \textbf{Fake-Mamba}—the first framework that re-architects Transformer/Conformer modules by substituting multi-head self-attention with bidirectional state-space modeling for speech
deepfake detection. Leveraging XLSR's rich linguistic representations, the proposed BiMamba encoders (TransBiMamba/ConBiMamba/PN-BiMamba) enable efficient spectral-temporal feature learning. The PN-BiMamba variant specifically employs Pre-LayerNorm stabilization and bidirectional feature fusion to localize subtle synthetic cues. The proposed framework is evaluated on three public benchmark databases (ASVspoof 2021 LA~\cite{Yamagishi2021ASVspoof}, ASVspoof 2021 DF~\cite{Yamagishi2021ASVspoof}, and In-the-Wild~\cite{ref14}), and results demonstrate the robustness over SOTA systems, while maintaining real-time inference across variable durations.

\section{Related Works}

\subsection{Mamba for Audio and Speech Processing}

Mamba \cite{ref26} has demonstrated Transformer-level performance across various time sequence-based modalities, including audio and speech, which are naturally sequential in waveform or spectrogram forms. Early works applied Mamba to single audio-related tasks such as audio representation learning~\cite{erol2024audio,yadav2024audio,shams2024ssamba}, polyphonic audio classification~\cite{lee2025deft}, and audio super-resolution~\cite{zhang2025vm}. This capability extends to speech processing, with applications in speech recognition~\cite{gao2024speech}, speech enhancement~\cite{chao2024investigation, 10890412}, speech separation~\cite{jiang2025dual}, and speaker-related tasks including speaker separation~\cite{avenstrup2025sepmamba}, speaker diarization~\cite{plaquet2025mamba}, target speaker extraction~\cite{fan2025improved} and speaker verification~\cite{xuanasv1,xuanasv2,xuanasv3,xuanasv4,xuan2024efficient}. However, the design of efficient Mamba architectures for real-time SDD remains underexplored.

\subsection{Mamba for Speech Deepfake Detection}

Despite the growing interest in Mamba-based models for audio and speech processing, the exploration of Mamba-based architectures in SDD remains limited. RawBMamba \cite{ref34} first introduced a bidirectional Mamba for SDD, utilizing SincNet and convolutional layers for short-range features alongside bidirectional Mamba blocks to capture long-range dependencies. A fusion module combines embeddings to integrate multi-scale information. To mitigate limited front-end representational capacity, XLSR-Mamba \cite{ref35} proposed a dual-column bidirectional Mamba leveraging self-supervised wav2vec 2.0. Critically, both frameworks rely on stacked pure Mamba blocks. This design constrains deep cross-dimensional interaction within the bidirectional architecture, making it harder to capture temporal-channel artifact cues essential for robust detection. To resolve this limitation, we propose three novel BiMamba variant refinements that enhance feature transformation and interaction mechanisms in bidirectional Mamba.

\section{Mathematical Foundations of the Mamba}

\begin{figure}[!ht]  
    \centering
    \includegraphics[width=0.75\columnwidth]{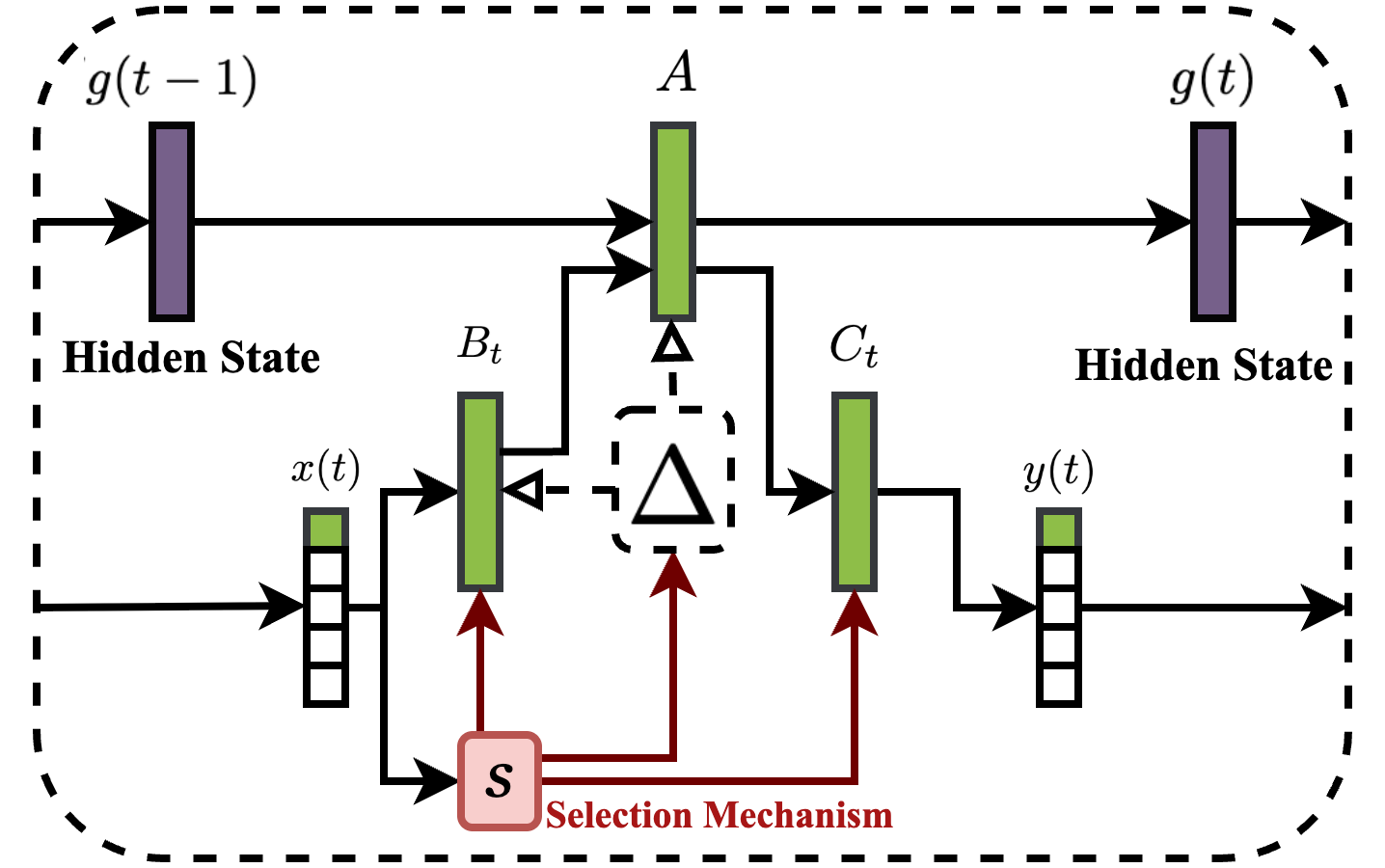}  
    \caption{An illustration of the State Space Model (SSM) with the Selection Mechanism (S). The model processes the input sequence through a core SSM module and dynamically selects relevant features using the Selection Mechanism (highlighted in red) before producing the final output.}
    \label{fig:ssm_selection}
\end{figure}

State space model ~\cite{hamilton1986state} formalizes a mathematically principled framework for sequence modeling, drawing inspiration from continuous systems that map a one-dimensional function or sequence $x(t) \in \mathbb{R}$ to an output $y(t) \in \mathbb{R}$ through a hidden state vector $g(t) \in \mathbb{R}^{N_M}$. This is achieved using evolution parameters $A \in \mathbb{R}^{N_M \times N_M}$ and projection parameters $B \in \mathbb{R}^{N_M \times 1}$ and $C \in \mathbb{R}^{1 \times N_M}$. Figure~\ref{fig:ssm_selection} illustrates this architecture. 

The continuous-time SSM is characterized by the following differential equations: 
\begin{equation}
g'(t) = Ag(t) + Bx(t),  y(t) = Cg(t)
\end{equation} 

Discretization is essential for deploying continuous-state space models in digital systems. The discrete version of the SSM includes a timescale parameter $\Delta$, which transforms the continuous-time matrices $A$ and $B$ to their discrete-time counterparts. The so-called zero-order hold (ZOH) method is commonly used for this transformation, defined as follows:

\begin{equation}
    A_d = \exp(\Delta A), 
\end{equation}
\begin{equation}
    B_d = (\Delta A)^{-1} (\exp(\Delta A) - I) \cdot \Delta B,
\end{equation}
where $I$ is the identity matrix and where $\exp(\cdot)$ is matrix exponential. After this discretization, the discrete-time SSM 
with step size $\Delta$ can be expressed as:
\begin{equation}
    g_t = A_d g_{t-1} + B_d x_t, g_t = C g_t. 
\end{equation}

\begin{figure*}[t] 
  \centering
  \includegraphics[width=0.90\textwidth]{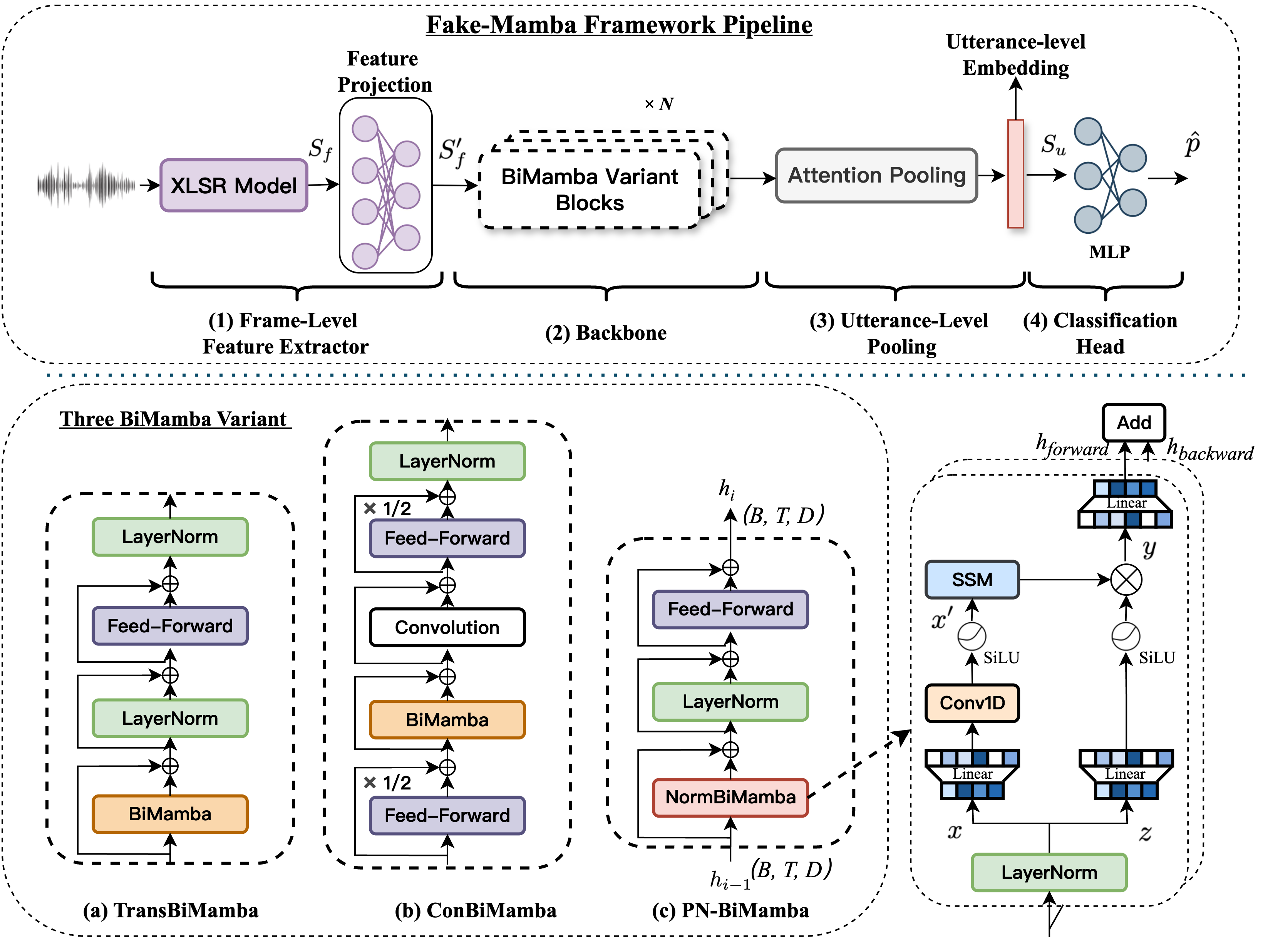} 
  \caption{Overview of the Fake-Mamba framework for SDD (top), with three different BiMamba variant configurations (bottom): (a) TransBiMamba; (b) ConBiMamba; and (c) PN-BiMamba; and four-stage processing Fake-Mamba framework pipeline:(1) Frame-Level feature extraction, (2) Backbone, (3) Utterance-level Pooling, and (4) Classification Head. The Classification Head is implemented as a multi-layer perceptron that maps features produced by the BiMamba variant blocks into whether the input speech is real or fake.}
  \label{fig:1}
\end{figure*}

To compute the output sequence $y_t$ efficiently, we use a global convolution operation. The output $y$ is obtained by convolving the input sequence $x$ with a structured convolutional kernel $K_d$ precomputed from the matrices $A_d$, $B_d$, and $C$:
\begin{equation}
    K_d = (CB_d, CA_d B_d, \dots, CA_d^{M-1} B_d),
\end{equation}
\begin{equation}
    y = x * K_d,
\end{equation}
where $M$ is the length of the input sequence $x$, and $K \in \mathbb{R}^M$ is the structured convolution kernel. The Mamba model enhances this framework by incorporating dynamic updates to the parameters $\Delta_t$, $A_t$, $B_t$, and $C_t$ based on the input $x_t$ at each timestep $t$. This makes the model input-selective and content-aware, allowing it to adjust to the specific characteristics of the input sequence dynamically. To efficiently handle these dynamic updates, Mamba employs a selective scan algorithm that recalculates the convolution dynamically, ensuring efficient and accurate sequence modeling.

\section{Fake-Mamba Framework}

In this section, we provide a detailed overview of Fake-Mamba, illustrated in Figure \ref{fig:1}. First, we describe how Fake-Mamba captures frame-level feature representation. Then, we present the implementation details of the encoder core—composed of three bidirectional Mamba variants; and the utterance-level pooling \& classification head.

\subsection{Frame-Level Feature Extractor: XLSR Model }

We employ XLSR~\cite{ref19}, one of the well-established foundational models, as the front-end feature extractor. 
It is a model for self-supervised cross-lingual speech representation learning based on wav2vec 2.0~\cite{baevski2020wav2vec}, pre-trained on 128 languages and about 436k hours of unlabeled speech data. XLSR shows strong performance in various downstream tasks, including speech recognition, translation, source tracing~\cite{xuan2025multilingual} and emotion recognition~\cite{learnnewlanguage, 10887615}. Having been pre-trained on large-scale realistic speech data in various languages, XLSR is efficient in capturing the essential characteristics of real human speech. Importantly, XLSR 
has been reported to outperform other 
foundation models in SDD~\cite{xin2022investigating}. For these reasons, XLSR forms an 
appropriate front-end 
to address SDD. 

During the supervised training phase, we fine-tune the parameters of XLSR 
concurrently with our backbone, utterance-level pooling and
classification head. 
Given an input audio, the corresponding frame-level feature representation \( S_f \in \mathbb{R}^{T \times C} \) is first extracted. Here, \( T \) and \( C \) are the number of time frames and channels, respectively. The feature representation is then fed to a learnable linear projection layer to reduce the channel dimensionality of features from \( C \) to \( D \). 
The reduced feature representation \( S_f^{\prime} \in \mathbb{R}^{T \times D} \) is used for the next steps. 

\subsection{Proposed Backbone} 
Existing bidirectional Mamba (BiMamba) architectures \cite{ref34,ref35} for the SDD suffer from representation bottlenecks due to pure block stacking. This design overlooks the dependencies between the temporal and channel dimensions, which are critical for detecting localized artifacts. To resolve this, inspired by \cite{transbimamba}, we propose three BiMamba variants for SDD:

\textbf{TransBiMamba} The structure of the TransBiMamba block is shown in Figure~\ref{fig:1}(a). We replace MHSA in the standard Transformer with BiMamba, termed TransBiMamba;

\textbf{ConBiMamba}
The structure of the ConBiMamba block is shown in Figure~\ref{fig:1}(b). We replace MHSA in the standard Conformer with BiMamba, termed ConBiMamba;

\textbf{PN-BiMamba} The structure of PN-BiMamba block as shown in Figure~\ref{fig:1}(c). Mathematically, for the input feature \( h_{i-1} \) to the \( i \)-th PN-BiMamba block, the output feature \( h_i \) is calculated by Equations (\ref{eq:1}) to (\ref{eq:11}):

\begin{equation}
\tilde{h}_{i-1} = \text{LayerNorm}( h_{i-1})
\label{eq:1}
\end{equation}
Then, $\tilde{h}_{i-1}$ is first projected onto two variables $x, z \in \mathbb{R}^{T \times E}$, which are used in the SSM module (Figure~\ref{fig:ssm_selection}). As shown in Equation \ref{eq:2}, the variable $x$ and the gating variable $z$ both depend on the input sequence $\tilde{h}_{i-1}$. Here, $E$ denotes the expanded state dimension, where $E \gg D$. 
\begin{align}
    x &= \text{Linear}_{\mathrm{x}}(\tilde{h}_{i-1}), & z &= \text{Linear}_{\mathrm{z}}(\tilde{h}_{i-1}).
    \label{eq:2}
\end{align}
The projected input sequence $x$ then passes through a 1D convolution and activation unit SiLU:

\begin{equation}
    x' = \text{SiLU}(\text{Conv1d}(x)),
    \label{eq:3}
\end{equation}
\begin{equation}
    y = \text{SSM}(x') \otimes \text{SiLU}(z),
    \label{eq:4}
\end{equation}
\begin{equation}
    h_{\mathrm{forward}} = \text{Linear}_{\mathrm{y}}(y), 
    \label{eq:5}
\end{equation}
\begin{equation}
h_{backward} = \text{Flip}(\text{Mamba}(\text{LayerNorm}(\text{Flip}({h}_{i-1}))))
\label{eq:6}
\end{equation}

Then, we add $h_{forward}$ and $h_{backward}$, thereby jointly encapsulating both left-to-right and right-to-left dependencies: 

\begin{equation}
h_{\mathrm{i-1}}' = h_{\mathrm{forward}} + h_{\mathrm{backward}}
\label{eq:8}
\end{equation}
\begin{equation}
h_{\mathrm{i-1}}'' = h_{\mathrm{i-1}}' + h_{\mathrm{i-1}}
\label{eq:8}
\end{equation}
\begin{equation}
h_{\mathrm{i-1}}''' = \text{LayerNorm}(h_{\mathrm{i-1}}'')
\label{eq:9}
\end{equation}
\begin{equation}
h_{\mathrm{i-1}}'''' = h_{\mathrm{i-1}}''' + h_{\mathrm{i-1}}''
\label{eq:10}
\end{equation}
\begin{equation}
h_{\mathrm{i}} = \text{FFN}(h_{\mathrm{i-1}}'''') + h_{\mathrm{i-1}}'''
\label{eq:11}
\end{equation}

where $x$, $z$, $\mathrm{ SiLU}(\cdot)$, $\mathrm{Mamba}(\cdot)$, $\mathrm{flip}(\cdot)$, $\mathrm{Conv1D}(\cdot)$, and $\otimes$ denote the input variable, input gating variable, sigmoid-weighted linear unit activation function, uni-directional Mamba block, flipping operation, one-dimensional convolution, and Hadamard product, respectively. Note that, \( h_{i-1} \in \mathbb{R}^{T \times D} \) and \( h_{i} \in \mathbb{R}^{T \times D} \), where $D$ is the PN-BiMamba encoder dimension.

\subsection{Utterance-level Pooling \& Classification Head}
The output of the encoder is fed to Linear Attention Pooling, obtaining utterance-level embedding \( S_u \in \mathbb{R}^D \), which is then forwarded to a multi-layer perceptron layer. The prediction \( \hat{p} \) is obtained from the logits for real and fake speech.

\section{Experimental Setup}

\input{Table/backbone}
\input{Table/sota}

\input{Table/ablation}
\input{Table/duration}

\subsection{Dataset}

We train all models on the ASVspoof 2019 LA (19LA) training data ~\cite{todsico2019asvspoof}, 
including $\sim$25k utterances and 6 spoofing attacks. 
To assess generalization performance, we perform three cross-dataset evaluations on the ASVspoof 2021 LA (21LA)~\cite{Yamagishi2021ASVspoof}, DF subsets (21DF)~\cite{Yamagishi2021ASVspoof}, and In-the-Wild (ITW) dataset ~\cite{ref14}. The first comprises $\sim$181k utterances with attacks similar to 19LA but with telephony encoding and transmission effects. The second contains $>$600k utterances and more than 100 attacks processed with various audio codecs, the third featuring $\sim$32k utterances collected under non-controlled conditions, making it a more challenging dataset.

\subsection{Data Augmentation}
To ensure comparability with ~\cite{ref18}, we use RawBoost ~\cite{tak2022rawboost} for data augmentation. Following~\cite{ref18,ref35,ref43,ref47}, we train two separate SDD models (with different RawBoost settings) for the 21LA and 21DF evaluation. In the 21LA, the SDD model was trained with RawBoost, combining linear and nonlinear convolutive noise with impulsive signal-dependent additive noise strategies. For the 21DF, stationary signal-independent additive noise with random coloration was introduced during training.

\subsection{Implementation Details}
All experiments were trained on Ubuntu 20.04.5 LTS, a single Tesla V100 32 GB GPU, with an identical random seed across all experiments. During training, audio inputs are fixed to 4.175s segments (66,800 samples) at a 16 kHz sampling rate.
The XLSR model was implemented using fairseq~\cite{ref48} and was jointly optimized with the backbone and classifier, without freezing any layers. The linear projection layer following the XLSR model has 144 output dimensions, which matches the dimension used for the BiMamba derivatives. Fake-Mamba(L) and Fake-Mamba(S) employ 7 and 4 PN-BiMamba blocks respectively, with a SSM expansion factor of 16. If not otherwise specified, “Fake-Mamba” refers to the (L) version.

Training was performed using the Adam optimizer with an initial learning rate of \(10^{-6}\), weight decay of \(10^{-4}\), and a batch size of 32. We trained with a WCE loss to account for class imbalance. The maximum number of training epochs is set to 100, with early stopping applied if the loss on the validation set does not decrease over 7 consecutive iterations. Based on the method given by the literature \cite{ref49,xuan2021}, the final model is derived by averaging the weights from the top 5 epochs with the lowest EERs on the development set.

\subsection{Evaluation Metrics}
We report both equal error rate (EER) and minimum normalized tandem detection cost function (min t-DCF). The former gauges real--fake discrimination, whereas the latter is the cost of automatic speaker verification subjected to spoofing. 
We further report 95\% parametric confidence intervals for EER following \cite{sholokhov2018semi}: 
\(\text{EER} \pm \sigma \cdot Z_{\alpha/2}\), where \(Z_{\alpha/2} = 1.96\), \(\sigma = 0.5 \sqrt{\text{EER}(1 - \text{EER}) \left({n_r + n_f})/({n_r n_f}\right)}\), and \(n_r\), \(n_f\) are the number of real and fake samples, respectively. Furthermore, we calculate the real-time factor (RTF) on the same GPU and device to evaluate the inference speed.

\section{Results}

\subsection{Framework with Different Mamba Derivatives}
\label{subsec:diff_mamba}
Table~\ref{tab:results} compares three BiMamba variants—ConBiMamba, TransBiMamba, and PN-BiMamba—as core components of Fake-Mamba. Across all benchmarks, PN-BiMamba consistently outperforms the others, while ConBiMamba yields the weakest results. This gap arises because PN-BiMamba’s parallel SSM paths and enhanced LayerNorm placement enable more effective fusion of temporal and channel information. Next, we compare PN-BiMamba-based Fake-Mamba(L).

\subsection{Comparison with SOTA Models}
The results of our proposed model are presented in Table~\ref{tab:main_results} along with a comparison to state-of-the-art models on the 21DF and ITW datasets. While Fake-Mamba(L) achieves comparable performance to the top-performing XLSR-DuaBiMamba~\cite{ref35} on 21LA, the EERs on 21DF (1.74\%) and ITW (5.85\%) indicate significant relative improvement over the previously best-reported XLSR-DuaBiMamba result by 7.45\% and 12.82\%, respectively. Moreover, Fake-Mamba(L) demonstrates significant improvements over XLSR-Conformer~\cite{ref18} across all three datasets, achieving performance gains of 29.71\%, 23.35\%, and 28.92\%, respectively. Additionally, our smaller Fake-Mamba(S) achieved a slight performance advantage on 21DF.

\begin{figure}[t]
    \centering
    \begin{subfigure}[b]{0.49\linewidth}
        \centering
        \begin{tikzpicture}
            \node[draw=black, very thick, inner sep=0pt] {
                \includegraphics[width=\linewidth]{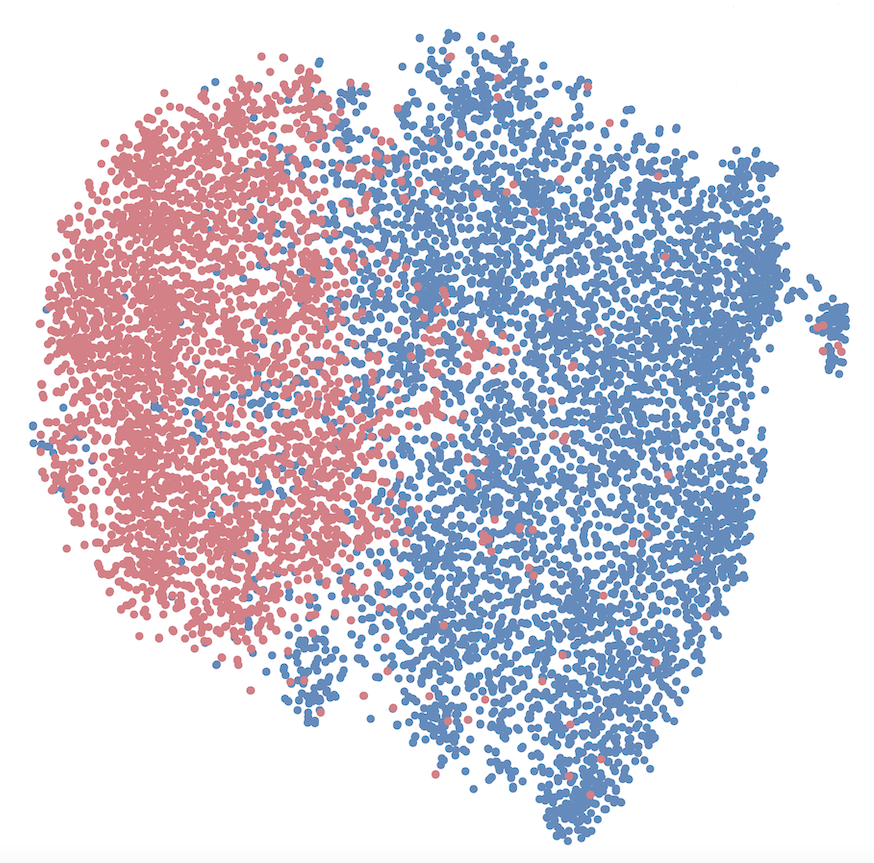}
            };
        \end{tikzpicture}
        \caption{XLSR-Conformer}
        \label{fig:tsne-conformer}
    \end{subfigure}
    \hfill
    \begin{subfigure}[b]{0.49\linewidth}
        \centering
        \begin{tikzpicture}
            \node[draw=black, very thick, inner sep=0pt] {
                \includegraphics[width=\linewidth]{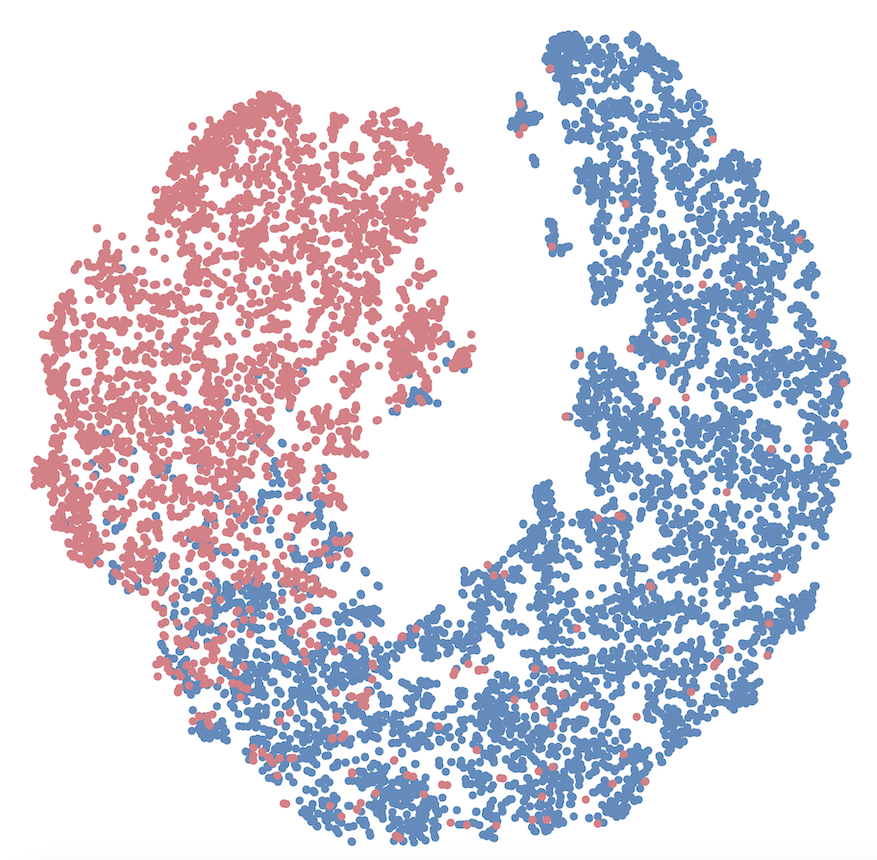}
            };
        \end{tikzpicture}
        \caption{Fake-Mamba}
        \label{fig:tsne-PN-BiMamba}
    \end{subfigure}
    
    \vspace{2mm}
    \noindent
    \begin{minipage}{\linewidth}
        \centering
        \textcolor{real_color}{\rule{1.2em}{1.2ex}}~Real\quad
        \textcolor{fake_color}{\rule{1.2em}{1.2ex}}~Fake
    \end{minipage}
    
    \caption{2D t-SNE visualization of In-the-Wild test set using model's utterance-level embeddings $S_u$.}
    \label{fig:tsne}
\end{figure}

\subsection{Ablation Study}
\subsubsection{Effectiveness of PN-BiMamba}
Given PN-BiMamba's superior performance in Section~\ref{subsec:diff_mamba}, we compare Fake-Mamba with PN-BiMamba blocks against Transformer and Conformer variants. Specifically, replacing PN-BiMamba blocks with Transformer or Conformer blocks and evaluating varying block counts on the ITW set reveals PN-BiMamba's significant advantages (Table~\ref{tab:ablation_results}). Notably, increasing block count fails to achieve comparable performance with Transformer and Conformer, demonstrating PN-BiMamba's robustness and generalization in SDD tasks.

\subsubsection{Component Ablation on PN-BiMamba}
We conducted an ablation study to quantify each PN-BiMamba component’s impact. Table~\ref{tab:ablation_results} shows that removing any single element substantially degrades performance. Omitting all three LayerNorm layers causes accuracy to fall to 62.6\% of the full-model result, underscoring their role in stabilizing training. Similarly, dropping the FFN reduces performance by 23.9\%, removing the bidirectional structure incurs a 35.2\% loss, and eliminating linear attention pooling leads to a 26.0\% decrease.


\subsection{Effect of Utterance Duration on Detection Performance}
Table~\ref{tab:2b} reports EERs for Fake-Mamba across different utterance lengths. Increasing the number of BiMamba blocks from 4 to 7 consistently lowers error rates for all variants, showing that deeper encoders capture more robust temporal patterns, especially in short clips. 

For very short utterances (under 3 seconds), ConBiMamba with seven blocks achieves the lowest EER at 9.44\%, slightly outperforming PN-BiMamba (9.75\%). As clip duration increases (3–4 s, 4–5 s, 5–6 s, and over 6 s), PN-BiMamba with seven blocks consistently delivers the best results, demonstrating the strongest generalization across utterance lengths. In comparison, ConBiMamba and TransBiMamba exhibit higher EERs in these mid-to-long duration ranges.

\subsection{Visualization}
Figure~\ref{fig:tsne} additionally visualizes utterance-level embeddings $S_u$ from XLSR-Conformer and Fake-Mamba(L) on the most demanding ITW dataset using t-SNE~\cite{ref51}. 
The embeddings from XLSR-Conformer exhibit substantial overlap between real and deepfake classes, while those from Fake-Mamba(L) display clearer separation. 

\subsection{Inference Speed}

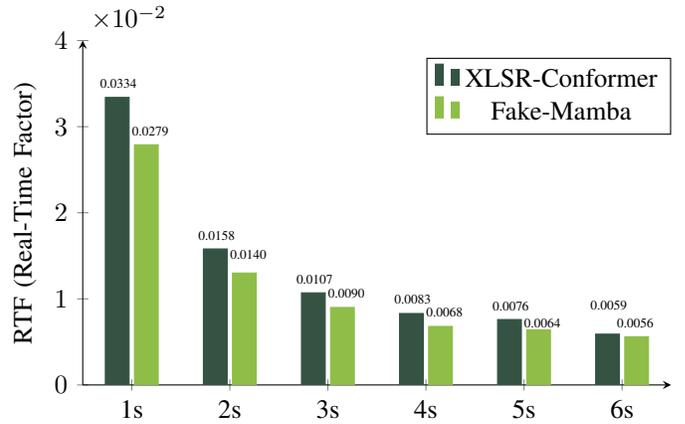
\begin{figure}[t]
    \centering
    \begin{tikzpicture}
        \begin{axis}[
            ybar,
            bar width=9pt,
            width=8.1cm,
            height=5.4cm,
            scale=1.2,
            every axis/.append style={transform shape},
            legend style={at={(1.00,0.95)}, anchor=north east, legend columns=1},
            symbolic x coords={1s,2s,3s,4s,5s,6s},
            xtick=data,
            ymin=0,
            ymax=0.04,
            ylabel={RTF (Real-Time Factor)},
            grid=none,
            axis lines=left,
            enlarge x limits={rel=0.1},
            ytick scale label code/.code={$\times 10^{#1}$}
            ]
            \addplot[color=dark_green, fill=dark_green] 
                coordinates {
                    (1s,0.0334) 
                    (2s,0.0158) 
                    (3s,0.0107) 
                    (4s,0.0083)
                    (5s,0.0076)
                    (6s,0.0059)};  
                    
            \addplot[color=light_green, fill=light_green] 
                coordinates {
                    (1s,0.0279) 
                    (2s,0.0130) 
                    (3s,0.0090) 
                    (4s,0.0068)
                    (5s,0.0064)
                    (6s,0.0056)};  
                    
            \legend{XLSR-Conformer, Fake-Mamba}

            \node[above, font=\tiny, color=black, xshift=-5pt] at (axis cs:1s,0.0334) {0.0334};
            \node[above, font=\tiny, color=black, xshift=-5pt] at (axis cs:2s,0.0158) {0.0158};
            \node[above, font=\tiny, color=black, xshift=-5pt] at (axis cs:3s,0.0107) {0.0107};
            \node[above, font=\tiny, color=black, xshift=-5pt] at (axis cs:4s,0.0083) {0.0083};
            \node[above, font=\tiny, color=black, xshift=-5pt] at (axis cs:5s,0.0076) {0.0076};
            \node[above, font=\tiny, color=black, xshift=-5pt] at (axis cs:6s,0.0076) {0.0059};

            \node[above, font=\tiny, color=black, xshift=7pt] at (axis cs:1s,0.0279) {0.0279};
            \node[above, font=\tiny, color=black, xshift=7pt] at (axis cs:2s,0.0135) {0.0140};
            \node[above, font=\tiny, color=black, xshift=7pt] at (axis cs:3s,0.0090) {0.0090};
            \node[above, font=\tiny, color=black, xshift=7pt] at (axis cs:4s,0.0068) {0.0068};
            \node[above, font=\tiny, color=black, xshift=7pt] at (axis cs:5s,0.0056) {0.0064};
            \node[above, font=\tiny, color=black, xshift=5pt] at (axis cs:6s,0.0056) {0.0056};
        \end{axis}
    \end{tikzpicture}
    \caption{Inference speed comparison (real-time factor) between Fake-Mamba(L) and XLSR-Conformer [39] across utterance durations from 1 to 6 seconds.}
    \label{fig:speed}
\end{figure}

Figure~\ref{fig:speed} compares the inference speed of Fake-Mamba(L) with the latest XLSR-Conformer~\cite{ref18} model in terms of average Real-Time Factor (RTF) over 100 runs on the same hardware. To mitigate performance fluctuations caused by cold-start effects, a GPU warm-up phase is conducted before formal measurements, ensuring the hardware reaches thermal equilibrium and a stable performance state~\cite{cui2025unlocking}. Fake-Mamba(L) provides consistently lower RTFs for all the utterance durations, indicating its efficiency. These results demonstrate the benefits of the hardware-friendly design of the proposed Fake-Mamba. This makes it particularly attractive for real-time anti-spoofing applications where processing efficiency is critical.

\section{Conclusion}

We introduced a novel audio deepfake detector, 
Fake-Mamba, which integrates an XLSR front-end with a new PN-BiMamba architecture, is designed to adaptively capture long-range dependencies. 
Our experiments indicate that Fake-Mamba outperforms other state-of-the-art models both on the ASVspoof 2021 evaluation set and on the more challenging In-the-Wild dataset. Additionally, Fake-Mamba exhibits faster inference, making it particularly appealing in real-time applications (such as call centers, teleconferencing, and Internet audio streaming services). All in all, our results indicate that Mamba-based architectures are viable replacements for Transformers and Conformers. Our planned future work will explore its potential in source tracing tasks.

\bibliographystyle{IEEEtran}
\bibliography{mybib}

\end{document}

%% file: Table/backbone.tex
\begin{table}[t]
    \centering
    \caption{Experimental results of BiMamba derivatives on ASVspoof and In-the-Wild datasets. Best results are in bold. 95\% parametric confidence intervals for EER are shown in parentheses.}
    \label{tab:results}
    \setlength{\tabcolsep}{4pt}
    \begin{tabular}{l c c c c}
        \Xhline{1.2pt}
        & \multicolumn{2}{c}{\textbf{21LA}} & \multicolumn{1}{c}{\textbf{21DF}} & \multicolumn{1}{c}{\textbf{ITW}} \\
        \cmidrule(lr){2-3} \cmidrule(lr){4-4} \cmidrule(lr){5-5}
        \textbf{Backbone} & \textbf{EER (\%) $\downarrow$} & \textbf{min-tDCF} $\downarrow$ & \textbf{EER (\%)} $\downarrow$ & \textbf{EER (\%)} $\downarrow$\\
        \midrule
        TransBiMamba & 1.25 & 0.2180 & 2.55 & 7.59 \\
        \quad & ({\scriptsize $\pm0.0943$}) &  & ({\scriptsize $\pm0.1285$}) & ({\scriptsize $\pm0.3012$}) \\
        \midrule 
        
        ConBiMamba & 1.80 & 0.2322 & 2.50 & 8.11 \\
        \quad & ({\scriptsize $\pm0.1128$}) &  & ({\scriptsize $\pm0.1272$}) & ({\scriptsize $\pm0.3104$}) \\
        \midrule

        \textbf{PN-BiMamba} & \textbf{0.97} & \textbf{0.2113} & \textbf{1.74} & \textbf{5.85} \\
        & ({\scriptsize $\pm0.0832$}) &  & ({\scriptsize $\pm0.1065$}) & ({\scriptsize $\pm0.2668$}) \\
        \Xhline{1.2pt}
    \end{tabular}
    \vspace{-4mm}
\end{table}

%% file: Table/sota.tex
\begin{table*}[ht] 
\centering 
\caption{Comparison with SOTA single systems. Results of models trained on ASVspoof19 and tested on the 21LA, 21DF, and In-the-Wild datasets. (S) and (L) represent 4 and 7 PN-BiMamba blocks, respectively. (* means our reproduced result; Bold denotes the best results)} 
\label{tab:main_results} 
\small 
\renewcommand{\arraystretch}{1.02} 
\begin{adjustbox}{width=0.85\textwidth,center} 
\begin{tabular}{lcccccc} 
\Xhline{1.2pt}
\textbf{Model} & \textbf{Params (M)} & \multicolumn{2}{c}{\textbf{ASVspoof 2021 LA}} & \textbf{ASVspoof 2021 DF} & \textbf{In-the-Wild} \\  
& & \textbf{EER}(\%) $\downarrow$ & \textbf{min-tDCF} $\downarrow$ & \textbf{EER}(\%) $\downarrow$ & \textbf{EER}(\%) $\downarrow$ \\  
\Xhline{0.8pt}
\noalign{\vskip 1.5pt} 

RawBMamba~\cite{ref34} & 0.719 &  3.28{\scriptsize $\pm0.1511$} & 0.2709 & 15.85{\scriptsize $\pm0.2975$} & -\\ 
\noalign{\vskip 1.2pt}

XLSR-Conformer~\cite{ref18} & 319.74 & 1.38{\scriptsize $\pm0.0978$} & 0.2216 & 2.27{\scriptsize $\pm0.1215$} & - \\ 
\noalign{\vskip 1.2pt}

XLSR-Conformer * & 319.74 & 1.40{\scriptsize $\pm0.0990$} & 0.2226 & 2.89{\scriptsize $\pm0.1370$} & 8.23{\scriptsize $\pm0.3127$}\\
\noalign{\vskip 1.2pt}

XLSR-Conformer+TCM~\cite{ref47} & 319.77 & 1.03{\scriptsize $\pm0.0849$} & 0.2130 & 2.06{\scriptsize $\pm0.1149$} & 7.79{\scriptsize $\pm0.3034$}\\ 
\noalign{\vskip 1.2pt}

XLSR-SLS~\cite{ref43} & 341.49 & 2.87{\scriptsize $\pm0.1403$} & - & 1.92{\scriptsize $\pm0.1104$} & 7.46{\scriptsize $\pm0.2960$} \\
\noalign{\vskip 1.2pt}

XLSR-DuaBiMamba~\cite{ref35} & 319.33 & \textbf{0.93{\scriptsize $\pm0.0807$}} & \textbf{0.2080} & 1.88{\scriptsize $\pm0.1086$} & 6.71{\scriptsize $\pm0.2795$} \\
\noalign{\vskip 1.5pt}

\midrule
\noalign{\vskip 1.5pt}

\textbf{Fake-Mamba (S)} & 318.79 & 1.19{\scriptsize $\pm0.0920$} & 0.2174 & 1.86{\scriptsize $\pm0.1100$} & 8.02{\scriptsize $\pm0.3088$} \\  
\noalign{\vskip 1.5pt}

\textbf{Fake-Mamba (L)} & 319.72 & 0.97{\scriptsize $\pm0.0832$} & 0.2113 & \textbf{1.74{\scriptsize $\pm0.1065$}} & \textbf{5.85{\scriptsize $\pm0.2668$}} \\  
\noalign{\vskip 1.0pt}

\Xhline{1.2pt}
\end{tabular} 
\end{adjustbox} 
\end{table*}

%% file: Table/ablation.tex
\begin{table}[t]
    \centering
    \caption{Ablation study results for Fake-Mamba (PN-BiMamba as backbone) on 21LA, 21DF, and In-the-Wild datasets. Best results are in bold. 95\% parametric confidence intervals for EER are shown in parentheses.}
    \label{tab:ablation_results}
    \small
    \setlength{\tabcolsep}{6pt}
    \renewcommand{\arraystretch}{1.02} 
    \begin{tabular}{@{}l*{3}{c}@{}}
        \Xhline{1.4pt}
        & \multicolumn{1}{c}{\textbf{21LA}} & \multicolumn{1}{c}{\textbf{21DF}} & \multicolumn{1}{c}{\textbf{ITW}} \\
        \cmidrule(lr){2-2} \cmidrule(lr){3-3} \cmidrule(lr){4-4}
        & \textbf{EER (\%)} & \textbf{EER (\%)} & \textbf{EER (\%)} \\
        \Xhline{0.5pt}
        \noalign{\vskip 1.2pt} 
        
        \textbf{Fake-Mamba} & \textbf{0.97} & \textbf{1.74} & \textbf{5.85} \\
        \quad &({\scriptsize $\pm0.0832$}) & ({\scriptsize $\pm0.1065$}) & ({\scriptsize $\pm0.2668$}) \\
        
        \noalign{\vskip 1.5pt} 
        \Xhline{0.8pt}
        \noalign{\vskip 2.0pt}
        \multicolumn{4}{c}{\textbf{Ablation 1: Fake-Mamba vs Conformer vs Transformer}} \\
        \noalign{\vskip 1.5pt}
        \Xhline{0.3pt}
        \noalign{\vskip 1.8pt}
        
        w Transformer(N=4) & 1.11 & 2.18 & 10.22 \\
        \quad &({\scriptsize $\pm0.0889$}) & ({\scriptsize $\pm0.1190$}) & ({\scriptsize $\pm0.3446$}) \\
        
        \noalign{\vskip 1.2pt}
        w Transformer(N=7) & 1.07 & 2.66 & 7.24 \\
        \quad &({\scriptsize $\pm0.0873$}) & ({\scriptsize $\pm0.1311$}) & ({\scriptsize $\pm0.2948$}) \\
        
        \noalign{\vskip 1.2pt}
        w Conformer(N=4) & 1.34 & 2.33 & 11.54 \\
        \quad &({\scriptsize $\pm0.0975$}) & ({\scriptsize $\pm0.1230$}) & ({\scriptsize $\pm0.3634$}) \\
        
        \noalign{\vskip 1.2pt}
        w Conformer(N=7) & 1.88 & 2.66 & 8.12 \\
        \quad &({\scriptsize $\pm0.1152$}) & ({\scriptsize $\pm0.1311$}) & ({\scriptsize $\pm0.3107$}) \\
        
        \noalign{\vskip 1.5pt}
        \Xhline{0.8pt}
        \noalign{\vskip 2.0pt}
        \multicolumn{4}{c}{\textbf{Ablation 2: PN-BiMamba}} \\
        \noalign{\vskip 1.5pt}
        \Xhline{0.3pt}
        \noalign{\vskip 1.8pt}
        
        w/o 3 Pre-LN & 3.00 & 4.13 & 9.51 \\
        \quad &({\scriptsize $\pm0.1448$}) & ({\scriptsize $\pm0.1622$}) & ({\scriptsize $\pm0.3341$}) \\
        
        \noalign{\vskip 1.2pt}
        w/o FFN & 1.07 & 2.47 & 7.25 \\
        \quad &({\scriptsize $\pm0.0873$}) & ({\scriptsize $\pm0.1265$}) & ({\scriptsize $\pm0.2945$}) \\
        
        \noalign{\vskip 1.2pt}
        w/o Bidirectional & 1.56 & 3.02 & 7.91 \\
        \quad &({\scriptsize $\pm0.1052$}) & ({\scriptsize $\pm0.1395$}) & ({\scriptsize $\pm0.3056$}) \\
        
        \noalign{\vskip 1.2pt}
        w/o Pooling & 2.19 & 2.52 & 7.37 \\
        \quad &({\scriptsize $\pm0.1242$}) & ({\scriptsize $\pm0.1277$}) & ({\scriptsize $\pm0.2973$}) \\
        
        \Xhline{1.4pt}
    \end{tabular}
\end{table}

%% file: Table/duration.tex
\begin{table}[t]
    \centering
    \caption{EER (\%) performance of Fake-Mamba with three BiMamba variants (TransBiMamba, ConBiMamba, and PN-BiMamba) across varying utterance duration ranges on the In-the-Wild test set. (Bold denotes the best results)}
    \label{tab:2b}
    \small
    \setlength{\tabcolsep}{6pt}
    \renewcommand{\arraystretch}{1.1}
    \begin{tabular}{@{}l*{6}{c}@{}}
        \Xhline{1.2pt}
        \noalign{\vskip 1.5pt}
        \textbf{Block} & \textbf{N} & \textbf{$<$3s} & \textbf{3-4s} & \textbf{4-5s} & \textbf{5-6s} & \textbf{$>$6s} \\
        \Xhline{0.9pt}
        \noalign{\vskip 1.8pt}
        \multirow{2}{*}{TransBiMamba} & 4  & 15.14 & 9.51 & 7.26 & 7.22 & 5.48 \\
                                     & 7  & 10.99 & 6.15 & 4.19 & 4.03 & 3.44 \\
        \midrule
        \multirow{2}{*}{ConBiMamba}   & 4  & 13.97 & 9.64 & 6.21 & 5.02 & 4.76 \\
                                     & 7  & \textbf{9.44} & 5.12 & 3.97 & 4.58 & 3.88 \\
        \midrule
        \multirow{3}{*}[0.55\normalbaselineskip]{\textbf{PN-BiMamba}} & 4  & 10.76 & 5.84 & 4.63 & 4.47 & 4.46 \\
                                     & 7  & 9.75 & \textbf{4.45} & \textbf{3.60} & \textbf{2.83} & \textbf{2.71} \\
        \Xhline{1.2pt}
    \end{tabular}
\end{table}

%% file: IEEE-conference-template-062824.bbl
\begin{thebibliography}{10}
\providecommand{\url}[1]{#1}
\csname url@samestyle\endcsname
\providecommand{\newblock}{\relax}
\providecommand{\bibinfo}[2]{#2}
\providecommand{\BIBentrySTDinterwordspacing}{\spaceskip=0pt\relax}
\providecommand{\BIBentryALTinterwordstretchfactor}{4}
\providecommand{\BIBentryALTinterwordspacing}{\spaceskip=\fontdimen2\font plus
\BIBentryALTinterwordstretchfactor\fontdimen3\font minus \fontdimen4\font\relax}
\providecommand{\BIBforeignlanguage}[2]{{%
\expandafter\ifx\csname l@#1\endcsname\relax
\typeout{** WARNING: IEEEtran.bst: No hyphenation pattern has been}%
\typeout{** loaded for the language `#1'. Using the pattern for}%
\typeout{** the default language instead.}%
\else
\language=\csname l@#1\endcsname
\fi
#2}}
\providecommand{\BIBdecl}{\relax}
\BIBdecl

\bibitem{ref1}
Y.~Jeon, Y.~Kim, and G.~G. Lee, ``Enhancing zero-shot multi-speaker tts with negated speaker representations,'' in \emph{Proceedings of the AAAI Conference on Artificial Intelligence}, vol.~38, no.~16, 2024, pp. 18\,336--18\,344.

\bibitem{breezyvoice}
\BIBentryALTinterwordspacing
C.-J. Hsu, Y.-C. Lin, C.-C. Lin, W.-C. Chen, H.~L. Chung, C.-A. Li, Y.-C. Chen, C.-Y. Yu, M.-J. Lee, C.-C. Chen, R.-H. Huang, H.~yi~Lee, and D.-S. Shiu, ``Breezyvoice: Adapting tts for taiwanese mandarin with enhanced polyphone disambiguation -- challenges and insights,'' 2025. [Online]. Available: \url{https://arxiv.org/abs/2501.17790}
\BIBentrySTDinterwordspacing

\bibitem{ref2}
Y.~K. Kan, K.~Xu, H.~Li \emph{et~al.}, ``Voicedefense: Protecting automatic speaker verification models against black-box adversarial attacks,'' in \emph{Proc. Interspeech 2024}, 2024, pp. 517--521.

\bibitem{ref3}
M.~Li, Y.~Ahmadiadli, and X.~P. Zhang, ``Audio anti-spoofing detection: A survey,'' \emph{arXiv preprint arXiv:2404.13914}, 2024.

\bibitem{Xuan2025}
\BIBentryALTinterwordspacing
X.~Xuan, K.~kui Sin, Y.~Zhou, and C.~Kit, ``Translaw: Benchmarking large language models in multi-agent simulation of the collaborative translation,'' 2025. [Online]. Available: \url{https://arxiv.org/abs/2507.00875}
\BIBentrySTDinterwordspacing

\bibitem{zhang1}
W.~Zhang and C.~Luo, ``Ge-gnn: Gated edge-augmented graph neural network for fraud detection,'' \emph{IEEE Transactions on Big Data}, vol.~11, no.~4, pp. 1664--1676, 2025.

\bibitem{ding1}
B.~Ding, R.~Han, Z.~Ma, and X.~Xuan, ``Crowd density estimation based on multi-level attention maps,'' in \emph{2021 IEEE 5th Information Technology,Networking,Electronic and Automation Control Conference (ITNEC)}, vol.~5, 2021, pp. 1759--1765.

\bibitem{10832250}
J.~Du, I.-M. Lin, I.-H. Chiu, X.~Chen, H.~Wu, W.~Ren, Y.~Tsao, H.-Y. Lee, and J.-S.~R. Jang, ``Dfadd: The diffusion and flow-matching based audio deepfake dataset,'' in \emph{2024 IEEE Spoken Language Technology Workshop (SLT)}, 2024, pp. 921--928.

\bibitem{codecfake-omni}
\BIBentryALTinterwordspacing
J.~Du, X.~Chen, H.~Wu, L.~Zhang, I.-M. Lin, I.-H. Chiu, W.~Ren, Y.~Tseng, Y.~Tsao, J.-S.~R. Jang, and H.~yi~Lee, ``Codecfake-omni: A large-scale codec-based deepfake speech dataset,'' \emph{CoRR}, vol. abs/2501.08238, January 2025. [Online]. Available: \url{https://doi.org/10.48550/arXiv.2501.08238}
\BIBentrySTDinterwordspacing

\bibitem{ref21}
A.~Gulati, J.~Qin, C.~C. Chiu \emph{et~al.}, ``Conformer: Convolution-augmented transformer for speech recognition,'' pp. 5036--5040, 2020.

\bibitem{xuan2024conformer}
X.~Xuan, R.~Han, and J.~Gao, ``Conformer-based speaker recognition model for real-time multi-scenarios,'' \emph{Computer Engineering and Applications}, vol.~60, no.~7, pp. 147--156, 2024.

\bibitem{ref17}
H.~Shin, J.~Heo, J.~Kim \emph{et~al.}, ``Hm-conformer: A conformer-based audio deepfake detection system with hierarchical pooling and multi-level classification token aggregation methods,'' in \emph{ICASSP 2024-2024 IEEE International Conference on Acoustics, Speech and Signal Processing (ICASSP)}.\hskip 1em plus 0.5em minus 0.4em\relax IEEE, 2024, pp. 10\,581--10\,585.

\bibitem{ref47}
D.~T. Truong, R.~Tao, T.~Nguyen \emph{et~al.}, ``Temporal-channel modeling in multi-head self-attention for synthetic speech detection,'' in \emph{Proceedings of Interspeech 2024}, 2024, pp. 537--541.

\bibitem{vaswani2017attention}
\BIBentryALTinterwordspacing
A.~Vaswani, N.~Shazeer, N.~Parmar, J.~Uszkoreit, L.~Jones, A.~Gomez, L.~Kaiser, and I.~Polosukhin, ``Attention is all you need,'' in \emph{Advances in Neural Information Processing Systems}, vol.~30.\hskip 1em plus 0.5em minus 0.4em\relax NeurIPS, 2017. [Online]. Available: \url{https://arxiv.org/abs/1706.03762}
\BIBentrySTDinterwordspacing

\bibitem{ref22}
J.~Yang, R.~K. Das, and H.~Li, ``Significance of subband features for synthetic speech detection,'' \emph{IEEE Transactions on Information Forensics and Security}, vol.~15, pp. 2160--2170, 2019.

\bibitem{ref23}
K.~Sriskandaraja, V.~Sethu, P.~N. Le \emph{et~al.}, ``Investigation of sub-band discriminative information between spoofed and genuine speech,'' in \emph{Interspeech}, 2016, pp. 1710--1714.

\bibitem{zhang2}
\BIBentryALTinterwordspacing
W.~Zhang, D.~Xu, X.~Xuan, L.~Jiang, G.~Yao, R.~Han, X.~Lang, and C.~Luo, ``Addressing noise and stochasticity in fraud detection for service networks,'' 2025. [Online]. Available: \url{https://arxiv.org/abs/2505.00946}
\BIBentrySTDinterwordspacing

\bibitem{zhang3}
W.~Zhang, D.~Xu, G.~Yao, X.~Lin, R.~Guan, C.~Du, R.~Han, X.~Xuan, and C.~Luo, ``Frect: Frequency-augmented convolutional transformer for robust time series anomaly detection,'' in \emph{Advanced Intelligent Computing Technology and Applications}, D.-S. Huang, W.~Chen, Y.~Pan, and H.~Chen, Eds.\hskip 1em plus 0.5em minus 0.4em\relax Singapore: Springer Nature Singapore, 2025, pp. 15--26.

\bibitem{zhang4}
\BIBentryALTinterwordspacing
W.~Zhang and C.~Luo, ``Decomposition-based multi-scale transformer framework for time series anomaly detection,'' \emph{Neural Networks}, vol. 187, p. 107399, 2025. [Online]. Available: \url{https://www.sciencedirect.com/science/article/pii/S0893608025002783}
\BIBentrySTDinterwordspacing

\bibitem{linprime}
Z.~Lin, J.~Wang, R.~Li, F.~Shen, and X.~Xuan, ``Primek-net: Multi-scale spectral learning via group prime-kernel convolutional neural networks for single channel speech enhancement,'' in \emph{ICASSP 2025 - 2025 IEEE International Conference on Acoustics, Speech and Signal Processing (ICASSP)}, 2025, pp. 1--5.

\bibitem{wu-etal-2024-codec}
\BIBentryALTinterwordspacing
H.~Wu, H.-L. Chung, Y.-C. Lin, Y.-K. Wu, X.~Chen, Y.-C. Pai, H.-H. Wang, K.-W. Chang, A.~Liu, and H.-y. Lee, ``Codec-{SUPERB}: An in-depth analysis of sound codec models,'' in \emph{Findings of the Association for Computational Linguistics: ACL 2024}, L.-W. Ku, A.~Martins, and V.~Srikumar, Eds.\hskip 1em plus 0.5em minus 0.4em\relax Bangkok, Thailand: Association for Computational Linguistics, Aug. 2024, p. 10330–10348. [Online]. Available: \url{https://aclanthology.org/2024.findings-acl.616/}
\BIBentrySTDinterwordspacing

\bibitem{10849259}
W.~Ren, Y.-C. Lin, H.-C. Chou, H.~Wu, Y.-C. Wu, C.-C. Lee, H.-Y. Lee, H.-M. Wang, and Y.~Tsao, ``Emo-codec: An in-depth look at emotion preservation capacity of legacy and neural codec models with subjective and objective evaluations,'' in \emph{2024 Asia Pacific Signal and Information Processing Association Annual Summit and Conference (APSIPA ASC)}, 2024, pp. 1--6.

\bibitem{ref26}
\BIBentryALTinterwordspacing
A.~Gu and T.~Dao, ``Mamba: Linear-time sequence modeling with selective state spaces,'' in \emph{First Conference on Language Modeling}, 2024. [Online]. Available: \url{https://openreview.net/forum?id=tEYskw1VY2}
\BIBentrySTDinterwordspacing

\bibitem{zhao2025cobra}
H.~Zhao, M.~Zhang, W.~Zhao, and et~al., ``{Cobra: Extending Mamba to Multi-Modal Large Language Model for Efficient Inference},'' in \emph{Proceedings of the AAAI Conference on Artificial Intelligence}, vol.~39, no.~10, 2025, pp. 10\,421--10\,429.

\bibitem{lenz2025jamba}
B.~Lenz, O.~Lieber, A.~Arazi, and et~al., ``{Jamba: Hybrid Transformer-Mamba Language Models},'' in \emph{The Thirteenth International Conference on Learning Representations}, 2025.

\bibitem{ref28}
R.~Waleffe, W.~Byeon, D.~Riach \emph{et~al.}, ``An empirical study of mamba-based language models,'' \emph{arXiv preprint arXiv:2406.07887}, 2024.

\bibitem{ref29}
L.~Zhu, B.~Liao, Q.~Zhang \emph{et~al.}, ``Vision mamba: Efficient visual representation learning with bidirectional state space model,'' in \emph{Forty-first International Conference on Machine Learning}, 2024.

\bibitem{wang2025mamba}
Z.~Wang, F.~Kong, S.~Feng, and et~al., ``{Is Mamba Effective for Time Series Forecasting?}'' \emph{Neurocomputing}, vol. 619, p. 129178, 2025.

\bibitem{li2024cmmamba}
Q.~Li, J.~Qin, D.~Cui, and et~al., ``{CMMamba: Channel Mixing Mamba for Time Series Forecasting},'' \emph{Journal of Big Data}, vol.~11, no.~1, p. 153, 2024.

\bibitem{Yamagishi2021ASVspoof}
J.~Yamagishi, X.~Wang, M.~Todisco \emph{et~al.}, ``Asvspoof 2021: Accelerating progress in spoofed and deepfake speech detection,'' in \emph{ASVspoof 2021 Workshop - Automatic Speaker Verification and Spoofing Countermeasures Challenge}, 2021.

\bibitem{ref14}
{Nicolas Müller and Pavel Czempin and Franziska Diekmann and Adam Froghyar and Konstantin Böttinger}, ``{Does Audio Deepfake Detection Generalize?}'' in \emph{{Interspeech 2022}}, {2022}, pp. {2783--2787}.

\bibitem{erol2024audio}
M.~H. Erol, A.~Senocak, J.~Feng, and et~al., ``{Audio Mamba: Bidirectional State Space Model for Audio Representation Learning},'' \emph{IEEE Signal Processing Letters}, 2024.

\bibitem{yadav2024audio}
S.~Yadav and Z.-H. Tan, ``{Audio Mamba: Selective State Spaces for Self-Supervised Audio Representations},'' in \emph{Proceedings of the 25th International Conference on Speech Communication and Technology (Interspeech 2024)}, 2024, pp. 552--556.

\bibitem{shams2024ssamba}
S.~Shams, S.~S. Dindar, X.~Jiang, and et~al., ``{SSAMBA: Self-Supervised Audio Representation Learning with Mamba State Space Model},'' in \emph{2024 IEEE Spoken Language Technology Workshop (SLT)}.\hskip 1em plus 0.5em minus 0.4em\relax IEEE, 2024, pp. 1053--1059.

\bibitem{lee2025deft}
D.~Lee and J.~W. Choi, ``{DeFT-Mamba: Universal Multichannel Sound Separation and Polyphonic Audio Classification},'' in \emph{ICASSP 2025 - 2025 IEEE International Conference on Acoustics, Speech and Signal Processing (ICASSP)}.\hskip 1em plus 0.5em minus 0.4em\relax IEEE, 2025, pp. 1--5.

\bibitem{zhang2025vm}
T.~Zhang and S.~Ruan, ``{VM-ASR: A Lightweight Dual-Stream U-Net Model for Efficient Audio Super-Resolution},'' \emph{IEEE Transactions on Audio, Speech and Language Processing}, 2025.

\bibitem{gao2024speech}
X.~Gao and N.~F. Chen, ``{Speech-Mamba: Long-Context Speech Recognition with Selective State Space Models},'' in \emph{2024 IEEE Spoken Language Technology Workshop (SLT)}.\hskip 1em plus 0.5em minus 0.4em\relax IEEE, 2024, pp. 1--8.

\bibitem{chao2024investigation}
R.~Chao, W.-H. Cheng, M.~La~Quatra, and et~al., ``An investigation of incorporating mamba for speech enhancement,'' in \emph{2024 IEEE Spoken Language Technology Workshop (SLT)}.\hskip 1em plus 0.5em minus 0.4em\relax IEEE, 2024, pp. 302--308.

\bibitem{10890412}
W.~Ren, H.~Wu \emph{et~al.}, ``Leveraging joint spectral and spatial learning with mamba for multichannel speech enhancement,'' in \emph{ICASSP 2025 - 2025 IEEE International Conference on Acoustics, Speech and Signal Processing (ICASSP)}, 2025, pp. 1--5.

\bibitem{jiang2025dual}
X.~Jiang, C.~Han, and N.~Mesgarani, ``{Dual-Path Mamba: Short and Long-Term Bidirectional Selective Structured State Space Models for Speech Separation},'' in \emph{ICASSP 2025 - 2025 IEEE International Conference on Acoustics, Speech and Signal Processing (ICASSP)}.\hskip 1em plus 0.5em minus 0.4em\relax IEEE, 2025, pp. 1--5.

\bibitem{avenstrup2025sepmamba}
T.~H. Avenstrup, B.~Elek, I.~L. M{\'a}di, and et~al., ``{SepMamba: State-Space Models for Speaker Separation Using Mamba},'' in \emph{ICASSP 2025 - 2025 IEEE International Conference on Acoustics, Speech and Signal Processing (ICASSP)}.\hskip 1em plus 0.5em minus 0.4em\relax IEEE, 2025, pp. 1--5.

\bibitem{plaquet2025mamba}
A.~Plaquet, N.~Tawara, M.~Delcroix, and et~al., ``{Mamba-Based Segmentation Model for Speaker Diarization},'' in \emph{ICASSP 2025 - 2025 IEEE International Conference on Acoustics, Speech and Signal Processing (ICASSP)}.\hskip 1em plus 0.5em minus 0.4em\relax IEEE, 2025, pp. 1--5.

\bibitem{fan2025improved}
C.~Fan, Y.~Gao, Z.~Pan, J.~Zhang, H.~Zhang, J.~Zhang, and Z.~Lv, ``Improved feature extraction network for neuro-oriented target speaker extraction,'' in \emph{ICASSP 2025 - 2025 IEEE International Conference on Acoustics, Speech and Signal Processing (ICASSP)}, 2025, pp. 1--5.

\bibitem{xuanasv1}
X.~Xuan, J.~Dong, and T.~Xuan, ``Research on front-end of asv system based on mel spectrum in noise scenario,'' in \emph{2022 IEEE 10th Joint International Information Technology and Artificial Intelligence Conference (ITAIC)}, vol.~10, 2022, pp. 2638--2642.

\bibitem{xuanasv2}
X.~Xuan and R.~Han, ``Research on acoustic feature extractor for automatic speaker verification systerm,'' in \emph{2022 IEEE 10th Joint International Information Technology and Artificial Intelligence Conference (ITAIC)}, vol.~10, 2022, pp. 2628--2633.

\bibitem{xuanasv3}
X.~Xuan, R.~Jin, T.~Xuan, G.~Du, and K.~Xuan, ``Multi-scene robust speaker verification system built on improved ecapa-tdnn,'' in \emph{2022 IEEE 6th Advanced Information Technology, Electronic and Automation Control Conference (IAEAC )}, 2022, pp. 1689--1693.

\bibitem{xuanasv4}
X.~Xuan, R.~Han, and B.~Ding, ``Research on speaker identification models based on cnn and additive angular margin loss,'' in \emph{2021 2nd International Conference on Electronics, Communications and Information Technology (CECIT)}, 2021, pp. 1046--1050.

\bibitem{xuan2024efficient}
X.~Xuan, Z.~Zhu, and C.~Kit, ``Efficient real-time multi-scenario speaker recognition with mel-spectrogram-based hybrid tdnn for edge system,'' in \emph{INTERSPEECH 2024-Young Female* Researchers in Speech Workshop (YFRSW 2024)}, 2024.

\bibitem{ref34}
Y.~Chen, J.~Yi, J.~Xue \emph{et~al.}, ``Rawbmamba: End-to-end bidirectional state space model for audio deepfake detection,'' pp. 2720--2724, 2024.

\bibitem{ref35}
Y.~Xiao and R.~K. Das, ``Xlsr-mamba: A dual-column bidirectional state space model for spoofing attack detection,'' \emph{IEEE Signal Processing Letters}, vol.~32, pp. 1276--1280, 2025.

\bibitem{hamilton1986state}
J.~D. Hamilton, ``State-space models,'' \emph{Handbook of Econometrics}, vol.~4, pp. 3039--3080, 1986.

\bibitem{ref19}
{Arun Babu and Changhan Wang and Andros Tjandra and Kushal Lakhotia and Qiantong Xu and Naman Goyal and Kritika Singh and Patrick {von Platen} and Yatharth Saraf and Juan Pino and Alexei Baevski and Alexis Conneau and Michael Auli}, ``{XLS-R: Self-supervised Cross-lingual Speech Representation Learning at Scale},'' in \emph{{Interspeech 2022}}, {2022}, pp. {2278--2282}.

\bibitem{baevski2020wav2vec}
A.~Baevski, Y.~Zhou, A.~Mohamed, and et~al., ``wav2vec 2.0: A framework for self-supervised learning of speech representations,'' \emph{Advances in Neural Information Processing Systems}, vol.~33, pp. 12\,449--12\,460, 2020.

\bibitem{xuan2025multilingual}
X.~Xuan, Y.~Xiao, R.~K. Das, and T.~Kinnunen, ``Multilingual source tracing of speech deepfakes: A first benchmark,'' \emph{arXiv preprint arXiv:2508.04143}, 2025.

\bibitem{learnnewlanguage}
\BIBentryALTinterwordspacing
S.-H. Wang, Z.-C. Chen, J.~Shi, M.-T. Chuang, G.-T. Lin, K.-P. Huang, D.~Harwath, S.-W. Li, and H.~yi~Lee, ``How to learn a new language? an efficient solution for self-supervised learning models unseen languages adaption in low-resource scenario,'' 2025. [Online]. Available: \url{https://arxiv.org/abs/2411.18217}
\BIBentrySTDinterwordspacing

\bibitem{10887615}
H.-C. Lin, Y.-C. Lin \emph{et~al.}, ``Improving speech emotion recognition in under-resourced languages via speech-to-speech translation with bootstrapping data selection,'' in \emph{ICASSP 2025 - 2025 IEEE International Conference on Acoustics, Speech and Signal Processing (ICASSP)}, 2025, pp. 1--5.

\bibitem{xin2022investigating}
X.~W., ``Investigating self-supervised front ends for speech spoofing countermeasures,'' in \emph{The Speaker and Language Recognition Workshop (Odyssey 2022)}, 2022, pp. 112--119.

\bibitem{transbimamba}
X.~Zhang, Q.~Zhang, H.~Liu, T.~Xiao, X.~Qian, B.~Ahmed, E.~Ambikairajah, H.~Li, and J.~Epps, ``Mamba in speech: Towards an alternative to self-attention,'' \emph{IEEE Transactions on Audio, Speech and Language Processing}, vol.~33, pp. 1933--1948, 2025.

\bibitem{ref18}
E.~Rosello, A.~Gomez-Alanis, A.~M. Gomez, and A.~Peinado, ``A conformer-based classifier for variable-length utterance processing in anti-spoofing,'' in \emph{Interspeech 2023}, 2023, pp. 5281--5285.

\bibitem{ref43}
Q.~Zhang, S.~Wen, and T.~Hu, ``{Audio deepfake detection with self-supervised XLS-R and SLS classifier},'' in \emph{Proceedings of the 32nd ACM International Conference on Multimedia}, 2024, pp. 6765--6773.

\bibitem{todsico2019asvspoof}
M.~Todisco, X.~Wang \emph{et~al.}, ``{ASVspoof 2019: Future horizons in spoofed and fake audio detection},'' pp. 1008--1012, 2019.

\bibitem{tak2022rawboost}
H.~Tak, M.~Kamble, J.~Patino \emph{et~al.}, ``Rawboost: A raw data boosting and augmentation method applied to automatic speaker verification anti-spoofing,'' in \emph{ICASSP 2022-2022 IEEE International Conference on Acoustics, Speech and Signal Processing (ICASSP)}.\hskip 1em plus 0.5em minus 0.4em\relax IEEE, 2022, pp. 6382--6386.

\bibitem{ref48}
C.~Wang, W.-N. Hsu, Y.~Adi, A.~Polyak, A.~Lee, P.-J. Chen, J.~Gu, and J.~Pino, ``fairseq s{\textasciicircum}2: A scalable and integrable speech synthesis toolkit,'' in \emph{Proceedings of the 2021 Conference on Empirical Methods in Natural Language Processing: System Demonstrations}, H.~Adel and S.~Shi, Eds.\hskip 1em plus 0.5em minus 0.4em\relax Online and Punta Cana, Dominican Republic: Association for Computational Linguistics, Nov. 2021, pp. 143--152.

\bibitem{ref49}
Y.~Gao, C.~Herold, Z.~Yang, and H.~Ney, ``Revisiting checkpoint averaging for neural machine translation,'' in \emph{Findings of the Association for Computational Linguistics: AACL-IJCNLP 2022}, Y.~He, H.~Ji, S.~Li, Y.~Liu, and C.-H. Chang, Eds.\hskip 1em plus 0.5em minus 0.4em\relax Association for Computational Linguistics, Nov. 2022, pp. 188--196.

\bibitem{xuan2021}
X.~Xuan, R.~Han, S.~Ji, and B.~Ding, ``Research on clothing image classification models based on cnn and transfer learning,'' in \emph{2021 IEEE 5th Advanced Information Technology, Electronic and Automation Control Conference (IAEAC)}, vol.~5, 2021, pp. 1461--1466.

\bibitem{sholokhov2018semi}
A.~Sholokhov, M.~Sahidullah, and T.~Kinnunen, ``Semi-supervised speech activity detection with an application to automatic speaker verification,'' \emph{Computer Speech \& Language}, vol.~47, pp. 132--156, 2018.

\bibitem{ref51}
S.~Arora, W.~Hu, and P.~K. Kothari, ``An analysis of the t-sne algorithm for data visualization,'' in \emph{Conference on Learning Theory}.\hskip 1em plus 0.5em minus 0.4em\relax PMLR, 2018, pp. 1455--1462.

\bibitem{cui2025unlocking}
A.~Cui, C.~Zhao, X.~Deng, G.~Jiang, Y.~Yang, G.~Yao, R.~Han, W.~Zhang, and X.~Xuan, ``Unlocking the full potential of separable convolutions on tensor cores,'' in \emph{International Conference on Intelligent Computing}.\hskip 1em plus 0.5em minus 0.4em\relax Springer, 2025, pp. 39--50.

\end{thebibliography}
